\documentclass{ws-p10x7}

\begin{document}

\title{Multi-Instanton Calculus in Supersymmetric Theories }

\author{F. Fucito}

\address{INFN, sez. di Roma 2, Via della Ricerca Scientifica, 00133
Roma, Italy\\E-mail: fucito@roma2.infn.it}

\twocolumn[\maketitle\abstract{
In this talk I review some recent results concerning multi-instanton
calculus in supersymmetric field theories. 
More in detail, I will show  how 
these computations can be efficiently performed using
the formalism of topological field theories.
}]

\section{Introduction}
Our understanding of the non--perturbative sector of field
and string theories has greatly progressed in recent
times. Recently, for the first time, the entire non--perturbative
contribution to the holomorphic part of the 
Wilsonian effective action was computed
for $N=2$ globally supersymmetric (SUSY) theories with gauge group $SU(2)$,
using ans\"atze dictated by physical intuitions \cite{sw}.
A few years later, 
a better understanding of non--perturbative configurations
in string theory led to the conjecture that 
certain IIB string theory correlators on an  
$AdS_5\times S^5$ background are related to 
Green's functions of composite operators of an 
$N=4$  $SU(N_c )$ Super Yang--Mills
(SYM) theory in four dimensions in the large $N_c$ limit
\cite{mal}. 
Although supported by many arguments,
these remarkable results remain conjectures and a clear mathematical 
proof seems to be out of reach at the moment. In our opinion
this state of affairs is mainly due to the lack of adequate computational 
tools
in the non--perturbative region. To the extent of our knowledge, the 
only way to perform computations in this regime 
in SUSY theories and from first principles 
 is via multi--instanton calculus.
Using this tool,  many partial checks 
have been performed on these conjectures,  
both in   $N=2$ and $N=4$ SUSY gauge 
theories \cite{fp}, \cite{DKM}, \cite{ft}, \cite{bkgr}, \cite{DKM3}.
The limits on these computations come from the exploding amount of algebraic
manipulations to be performed and from the lack
of an explicit parametrization of instantons of winding number greater
than two \cite{adhm}. 
In order to develop
new computational tools that might allow an extension 
to arbitrary winding  number, I revisit 
instanton computations for $N=2$ in the light of the topological theory 
built out of  $N=2$ SYM \cite{bftt},
{\it i.e.} the so--called Topological Yang--Mills 
theory (TYM) \cite{witten}.

\section{Topological Yang--Mills Theory}
Here I collect some results 
which will be relevant to our discussion.
I use the same notation of \cite{bftt} to which I refer the reader
for a detailed exposition of this material.
As it is well known \cite{bs}, after the twisting procedure, the
Lagrangian of $N=2$ SYM is invariant under
\begin{eqnarray}
\label{BRST} 
&&sA=\psi-Dc\ \
,\nonumber\\ &&s\psi=-[c,\psi]-D\phi\ \ ,\nonumber\\
&&s\phi=-[c,\phi]\ \ ,\nonumber\\ 
&&sc=-{1\over 2}[c,c]+\phi.
\end{eqnarray}
The BRST operator, $s$, defined 
in (\ref{BRST}) is
such that $s^2=0$ and when
the set of equations in (\ref{BRST}) is restricted
to the solutions of the
Euler--Lagrange classical equation ({\it the zero modes})
it gives the derivative on the space $M^{+}$.
In terms of the parameters of the ADHM construction,
these solutions look like
\begin{eqnarray}
A&=&U^\dagger d U,\\
\psi&=&U^{\dagger}{\cal M}f(d\Delta^{\dagger})U+
U^{\dagger}(d\Delta)f{\cal M}^{\dagger}U,\\ 
c&=&U^\dagger (s+C) U, \\
\phi&=& U^{\dagger}{\cal M} f {\cal M}^{\dagger}U+U^{\dagger}{\cal A} U.
\label{restoeq}
\end{eqnarray}
Plugging (\ref{restoeq}) into (\ref{BRST}) leads
to the action of the operator $s$ on the elements of the
ADHM construction
\begin{eqnarray}
{\cal M}&=&s\Delta+C\Delta={\cal S}\Delta\label{azzarolina1}\ \ ,\\
{\cal A}&=&s{\cal M}\Delta+C{\cal M}={\cal S}{\cal M}\label{azzarolina2}\ \ ,\\
s{\cal A}&=&-[C,{\cal A}]\label{azzarolina3}\ \ ,\\
sC&=&{\cal A}-C C\label{azzarolina4}\ \ ,
\end{eqnarray}
{\it i.e.} this is the realization of the BRST algebra on 
the instanton moduli space. $C$ is a connection I must introduce
to have a nilpotent $s$.

Let us see now how,
at the semi--classical level, any correlator which is expressed
as a polynomial in the fields, becomes after projection
onto the zero--mode subspace, a well--defined  
differential form on $M^+$ \cite{witten}.
Symbolically
\begin{equation}
\label{prescription}
\left< fields \right> = \int_{M^{+}}  \ 
\left[ (fields)\ e^{-S_{\rm TYM}} \right]_{zero-mode}
\ \ .
\end{equation}

A generic  function on the zero--mode subspace
will then have the expansion
\begin{eqnarray}
\label{inte1}
g( \widehat{\Delta}, \widehat{\cal M}) & = & 
g_{0} (\widehat{\Delta} )+ 
g_{i_1} (\widehat{\Delta} ) \widehat{\cal M}_{i_1} + 
\ldots 
\nonumber \\
&+&
{1\over p!}
g_{i_1 i_2 \ldots i_p} (\widehat{\Delta}) 
\widehat{\cal M}_{i_1}\widehat{\cal M}_{i_2}\cdots \widehat{\cal M}_{i_p}
\ \ ,
\end{eqnarray}
the coefficients of the expansion being totally antisymmetric in their indices.
Now  
$\widehat{\cal M}_{i}$'s and the $s\widehat{\Delta}_{i}$'s are related 
by a (moduli--dependent) linear transformation $K_{ij}$, 
which is completely known 
once the explicit expression for $C$ is 
plugged into the $\widehat{\cal M}_{i}$'s:
\begin{equation}
\label{inte2}
\widehat{\cal M}_{i} = K_{ij} ( \widehat{\Delta} ) s\widehat{\Delta}_{j}
\ \ .
\end{equation}
It then follows that 
\begin{eqnarray}
\label{noncera}
&&\widehat{\cal M}_{i_1}\widehat{\cal M}_{i_2}\cdots \widehat{\cal M}_{i_p}=
\nonumber \\
&=& \epsilon_{j_1 \ldots j_p} K_{i_1 j_1} K_{i_2 j_2} \cdots K_{i_p j_p} 
\ s^p \widehat{\Delta} =
\nonumber \\
&=&
\epsilon_{i_1 \ldots i_p} ({\rm det} K) \  s^p \widehat{\Delta}
\ \ ,
\end{eqnarray}
where $s^p \widehat{\Delta}\equiv 
s\widehat{\Delta}_1 \cdots s \widehat{\Delta}_p$.
I then conclude that
\begin{eqnarray}
&&\int_{M^+}g( \widehat{\Delta}, \widehat{\cal M})= 
{1\over p!} \int_{M^+} g_{i_1\ldots i_p} (\widehat{\Delta} ) 
\widehat{\cal M}_{i_1}\cdots \widehat{\cal M}_{i_p}
\nonumber \\
&=&
\int_{M^+}  s^p  \widehat{\Delta} \  |{\rm det} K| 
g_{1 2 \ldots p} (\widehat{\Delta} )
\ \ .
\label{finprescription}
\end{eqnarray}
The determinant of $K$ naturally stands out as 
{\it the instanton integration measure for $N=2$ SYM theories.}

\section{Conclusions}
In this talk i have argued that the results of multi-instanton 
calculus at the semiclassical level can be easily recovered in the formalism
of topological field theories. More benefits come from this reformulation
than what I have presented until now. It is for example possible to show that
correlators of the type of (\ref{prescription}) can be written as a total 
derivative on the moduli space. The only contributions to these quantities,
come from zero-size instantons. Given the peculiar properties of the ADHM
construction at the boundary of the moduli space, this might lead to recursion
relations among correlators computed at different winding number. To get to 
this conclusion, we probably have to better understand what is the 
geometrical role of
the action, projected on the subspace of the zero-modes.
The connection between moduli spaces of instantons and their
construction in terms of  D-branes of string theory can be very helpful in 
this respect.

\end{document}